\documentclass[preprint,aps,showpacs]{revtex4-1}
\usepackage{graphicx}
\begin{document}
\title{Mixtures of anisotropic and spherical colloids: Phase behavior, confinement, percolation phenomena and kinetics}
\author{T.~Schilling}
\affiliation{Theory of Soft Condensed Matter, Universit\'e du Luxembourg, L-1511 Luxembourg, Luxembourg}

\author{S.~Dorosz}
\affiliation{Theory of Soft Condensed Matter, Universit\'e du Luxembourg, L-1511 Luxembourg, Luxembourg}

\author{M.~Radu}
\affiliation{Theory of Soft Condensed Matter, Universit\'e du Luxembourg, L-1511 Luxembourg, Luxembourg}

\author{M.~Mathew}
\affiliation{KOMET Institut f\"ur Physik, Johannes Gutenberg-Universit\"at Mainz, Staudinger Weg 7, D-55099 Mainz, Germany}

\author{S.~Jungblut}
\affiliation{Fakult\"at f\"ur Physik, Universit\"at Wien, Boltzmanngasse 5, 1090 Wien, Austria}

\author{K.~Binder}
\affiliation{KOMET Institut f\"ur Physik, Johannes Gutenberg-Universit\"at Mainz, Staudinger Weg 7, D-55099 Mainz, Germany}

\begin{abstract}
Purely entropic systems such as suspensions of hard rods, platelets and spheres show rich phase behavior. Rods and platelets have successfully been used as models to predict the equilibrium properties of liquid crystals for several decades. Over the past years hard particle models have also been studied in the context of non-equilibrium statistical mechanics, in particular regarding the glass transition, jamming, sedimentation and crystallization. Recently suspensions of hard anisotropic particles also moved into the focus of materials scientists who work on conducting soft matter composites. An insulating polymer resin that is mixed with conductive filler particles becomes conductive when the filler percolates. In this context the mathematical topic of connectivity percolation finds an application in modern nano-technology.
In this article, we briefly review recent work on the phase behavior, confinement effects, percolation transition and phase transition kinetics in hard particle models. In the first part, we discuss the effects that particle anisotropy and depletion have on the percolation transition. In the second part, we present results on the kinetics of the liquid-to-crystal transition in suspensions of spheres and of ellipsoids.
\end{abstract} %end of abstract
\maketitle
\section{Introduction}
\label{intro}
Colloidal suspensions containing spherical \cite{1,2}, rod-like \cite{3,4} or plate-like \cite{5,6} particles in the size-range from tens of nanometers to a few micrometers are model systems for the study of cooperative phenomena in condensed matter (e.g. \cite{7,8,9,10}). Some of these systems also have influenced the design of novel functional materials in which fabrication consists of processes on the colloidal scale, e.g.~when electrically conducting carbon nanotubes are embedded in a polymeric matrix to obtain composite materials with favorable electrical properties, making them suitable for flat panel displays and photovoltaic devices \cite{11,12,13}. Related composite materials with either enhanced mechanical strength or improved thermal and/or electrical conductivity can also be obtained by mixing polymer resins with graphene sheets or graphite nanoplatelets \cite{14,15,16,17}. Such systems containing densely packed platelets show also a fascinating multitude of liquid crystalline ordering phenomena (nematic, smectic, columnar, etc.) \cite{18}. A very intriguing aspect is also the dynamics of such systems, which still poses many unsolved questions, such as the effect of solvent (or matrix-) mediated hydrodynamic interactions on the kinetics of phase transitions \cite{19,20}, or the interplay with glassy freezing \cite{21}.

In the present paper, we give a brief overview of selected results which elucidate some of the questions indicated above. In the first part, we study connectivity percolation \cite{22,23} in systems of hard rods \cite{24,25} that attract each other due to the presence of polymers in the suspension. (The polymers cause a depletion attraction analogous to the Asakura-Oosawa model \cite{26} for colloid-polymer mixtures \cite{27}). This system also exhibits a transition to a nematically ordered phase \cite{28}, but percolation occurs at much lower rod packing fractions \cite{24,25}, so there is no interplay of long-scale orientational ordering and percolation (Sec.~II). The situation, however, is different when one considers hard platelets \cite{29} (Sec.~III): then nematic order does have a pronounced effect on the percolation properties. In Sec.~IV we then shall consider the effect of hydrodynamic backflow on the kinetics of the formation of crystalline nuclei \cite{19}, restricting attention to the archetypical case of hard spheres, Sec.~V then considers crystallization kinetics in systems of hard ellipsoids \cite{21}, under conditions close to the glass transition \cite{30}, while Sec.VI presents a concluding discussion.
\section{Percolation Versus Phase Separation in Mixtures of Anisotropic and Spherical colloids}
In this section we consider a mixture of hard spherocylinders, of diameter $D$ (which we take as unit of length, $D=1$) and length $L$, with soft spheres (also of diameter $D=1)$. While the soft spheres can overlap with each other without any energy cost (they are thought to represent flexible polymer chains, which can interpenetrate each other in a solution under good solvent conditions), overlap of the spheres with the rods is strictly forbidden. Thus, these spheres create a depletion attraction between the rods, in analogy to the attraction between hard spheres in the standard Asakura-Oosawa (AO) model \cite{26,27}. The strength of this attraction is controlled by the packing fraction $\eta _s = (D^3\pi/6)(N_s/V)$ of the spheres (where $N_s$ is the number of spheres in the volume $V$). For computational purposes, it often is convenient to work with the fugacity $z_s$ of the spheres (which is an intensive thermodynamic variable, analogous to inverse temperature in a lattice gas model, for instance, while $\eta_s$ is the density of an extensive variable).

The phase diagram of this system has been studied both by free volume theory and by Monte Carlo simulations \cite{25,28,31,32,32a,32b}. It was found that a phase separation into a vapor-like phase (at small $\eta _r$) and a liquid-like phase (at larger $\eta_r$) occurs, see Fig.~1 which shows the phase diagram as a function of $\eta_s$ and $\eta_r=v_rN_r/V$, where the packing fraction of the rods $\eta_r$ is defined as the product of the number of rods ($N_r$) and the volume of one rod ($v_r=\pi D^2(2D+3L)/12)$ divided by the total volume. In Fig.~1, also the line of percolation transitions is included, that we shall discuss below.

Note that grand-canonical Monte Carlo simulations of the present model are not straightforward (under the conditions of interest the acceptance probability for a successful insertion of a spherocylinder would typically be less than 10$^{-6}$), so a ``cluster move'' needed to be developed \cite{25,31}, generalizing a related method for the standard AO model \cite{33}. The densities of the coexisting phases then are found by sampling their probability distribution (at various fixed values of $z_s$), applying successive umbrella sampling \cite{34} and the ``equal weight rule'' \cite{35}, for details see \cite{31}. (Also the modification of the phase diagram due to the confinement by planar walls has been studied \cite{25,32}, but this aspect shall stay outside of consideration here.)

\begin{figure}
\includegraphics[scale=0.37]{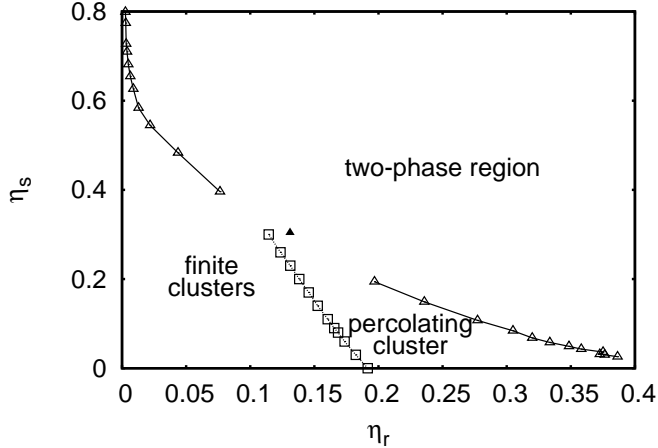}
\caption{\label{fig1} Phase diagram of the rod-sphere mixture in the plane of variables rod packing fraction $(\eta _r)$ and sphere packing fraction $(\eta _s)$, as obtained from grand-canonical Monte Carlo simulations \cite{25}, for the aspect ratio $L/D=3$. The critical point (full triangle) was located by finite size scaling methods, as described in \cite{28,31}. The open squares show the location of the percolation transition, which starts out at $\eta_r \simeq 0.19$ for $\eta_s =0$ and meets the vapor-like branch of the coexistence curve at $\eta_r \approx 0.11$. While to the left of this transition line only finite clusters of ``connected'' rods (see the main text) occur, an infinite (percolating) cluster is found to the right of this transition line. From \cite{25}.}
\end{figure}

We now turn to a discussion on how the percolation line in Fig.~1 has been located: recall that overlap of rods is strictly forbidden, and neither do configurations occur in which rods precisely touch each other. So a different connectivity criterion for the rods is required, which necessarily involves some arbitrariness. We have defined that two rods are connected if the distance $A$ of closest approach between their surfaces is less than $0.2D$ \cite{24,25} (or, equivalently, the line segments at the spherocylinder's axes approach closer than $1.2D$). This choice, while being arbitrary, reflects the fact that in conductivity experiments on carbon nanotube composites exact contact is not required either, because the electron transport between the tubes occurs mainly by tunneling. Using cubic simulation boxes of size $L_x \times L_x \times L_x$, one then asks whether or not a ``spanning cluster'' of rods (connected into itself via the periodic boundary conditions) occurs \cite{22,23}. Percolation defined in this way was tested every $20$ ``sweeps'' (trial move per particle) and thus the probability $\Pi$ that a spanning cluster occurs and the fraction $\phi$ of rods that it contains were sampled \cite{25}. Fig.~2 shows typical data for $\Pi$ for $L/D=4$, both for $\eta _s = 0$ and $\eta_s = 0.08$, and Fig.~2b shows a finite size scaling analysis \cite{36} of corresponding data for $\phi$.

\begin{figure}
\centering
\includegraphics[scale=0.37]{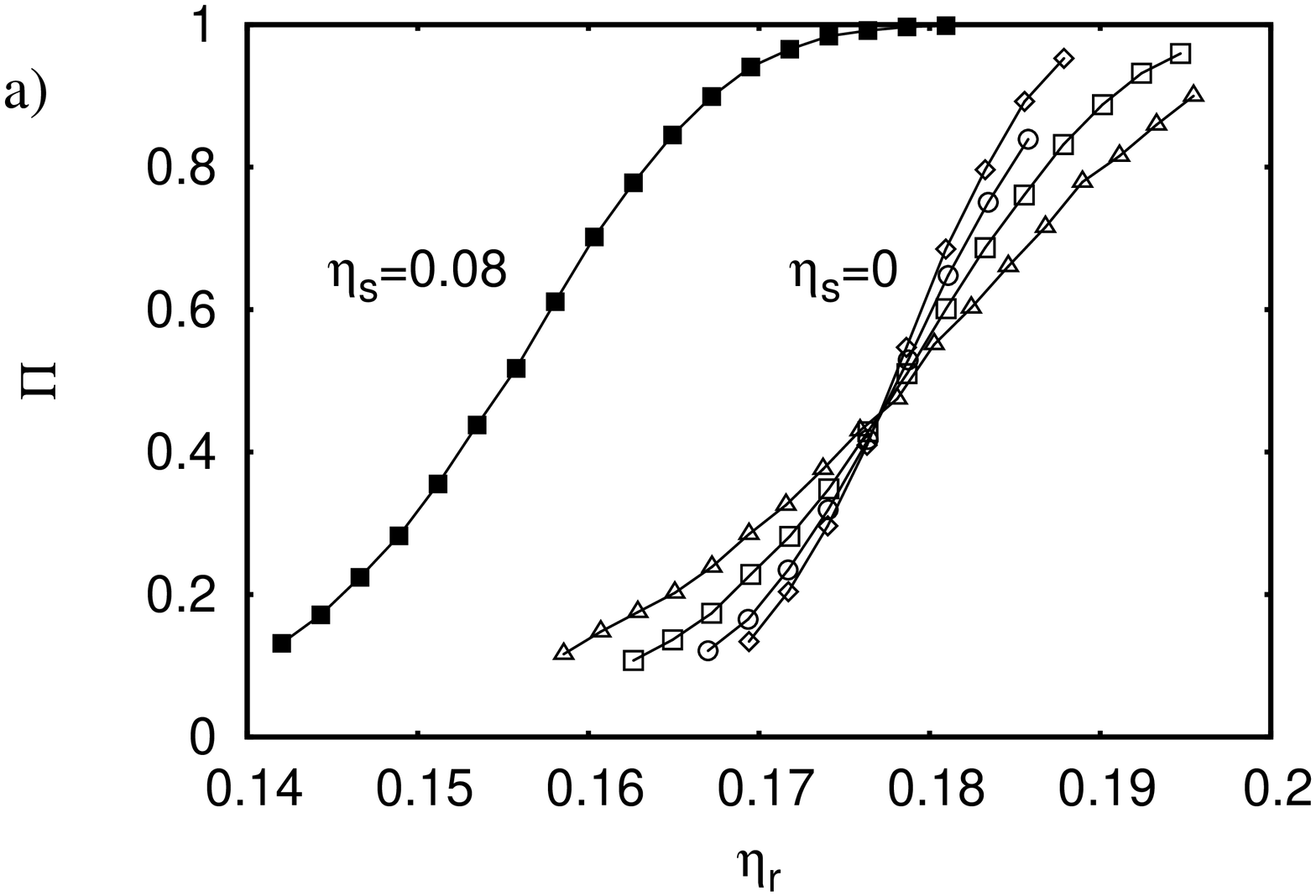}
\includegraphics[scale=0.37]{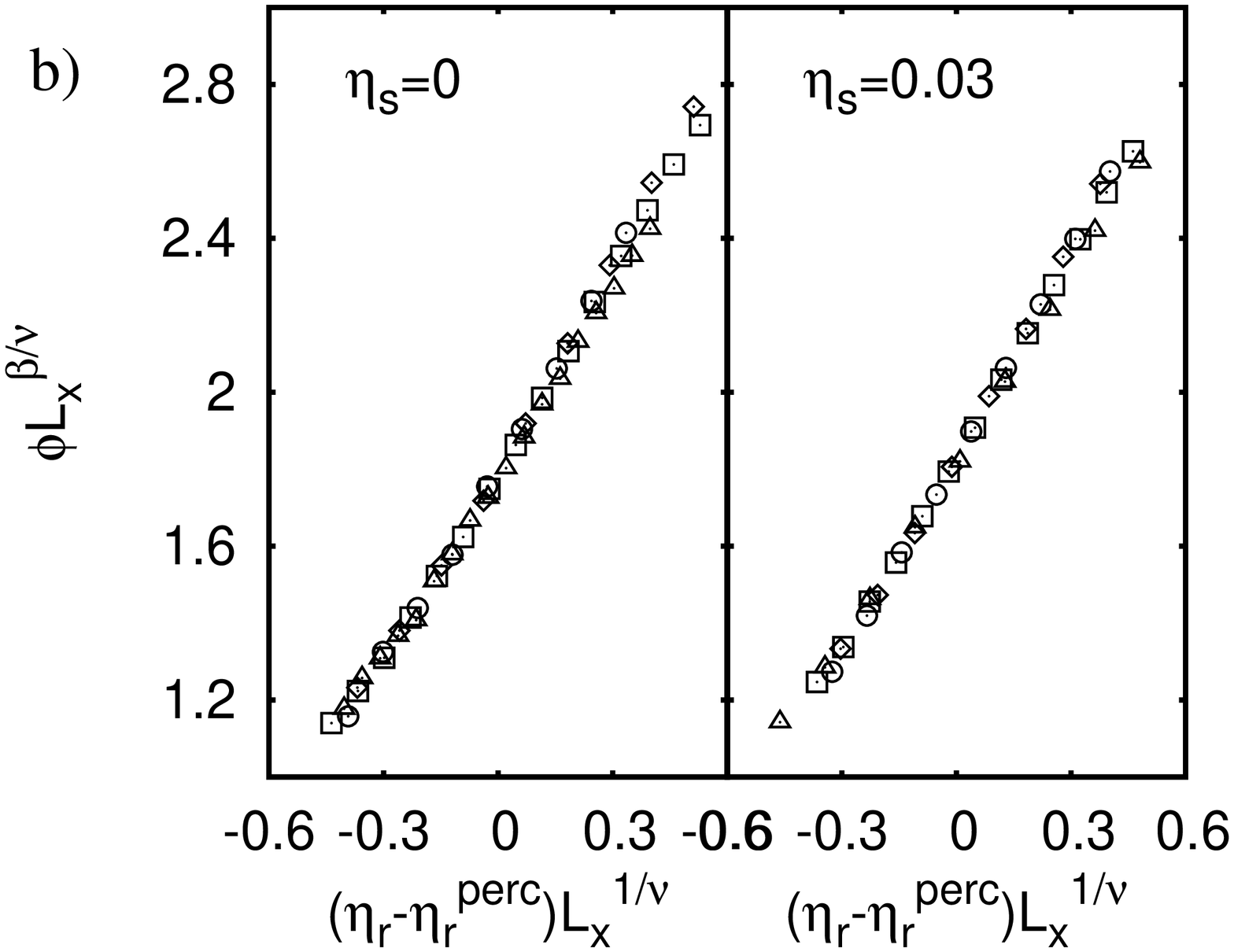}
\caption{\label{fig2} (a, top) Percolation probability $\Pi$ as a function of the rod packing fraction $\eta _r$ for an aspect ratio $L/D=4$ and two sphere packing fractions, $\eta_s=0$ (right) and $\eta_s =0.08$ (left). For $\eta_s=0$, several box linear dimensions are included: $L_x=15D$ (triangles), $20D$ (squares), $25D$ (circles), and $30D$ (diamonds). The filled squares for $\eta_s=0.08$ refer to a single choice $L_x=20D$. Curves are guides to the eye only. (b, bottom) Finite size scaling of the fraction $\phi$ of rods in the percolating cluster with the distance from the percolation threshold, $\eta_r-\eta_r^{\rm perc}$, i.e. $\phi L_x^{\beta /\nu}$ versus $(\eta_r-\eta_r^{\rm perc})L_x^{1/\nu}$, where $\nu \approx 0.88$ and $\beta \approx 0.41$ are the known \cite{22} critical exponent of random percolation. Two choices of $\eta_s$ are shown, as indicated. The meaning of the symbols is the same as in part (a). From \cite{25}.}
\end{figure}

Fig.~2a demonstrates that $\Pi$ for any $L_x$ is a nonsingular function of $\eta_r$, smoothly increasing from zero to unity as $\eta_r$ increases. But with increasing $L_x$, the curves get steeper, and there is a size-independent intersection point; i.e., $\Pi(\eta_r)$ converges to a step function at $\eta_r=\eta_r^{\rm perc}$ in the thermodynamic limit $L_x \rightarrow \infty$, and $\eta_r^{\rm perc}$ can be accurately estimated from the intersection point \cite{22}. Fig.~2b demonstrates that the critical behavior of this percolation transition still falls in the class of random percolation, despite the correlations between the rods that are present, in particular when $\eta_s >0$. This fact, of course, is expected, since the range of these correlations is finite. Note that the percolation transition line in Fig.~1 does not reach the critical point, but rather meets the coexistence curve on the vapor-like side, similar as in the lattice gas model \cite{36,37}, see Fig.~3. We see that both in the lattice gas model and in the rod-sphere mixture particle-particle attraction makes percolation easier, this effect is even more pronounced for the rod-sphere mixture, where the correlated percolation transition line starts out to the right of the critical rod packing fraction at $z_s=0$ (or $\eta_s = 0$, respectively), but bends over to the left side. Of course, the precise location of this line must depend somewhat on the arbitrary parameter $A$, but we expect that the general features of these phenomena are quite robust.

\begin{figure}
\centering
\includegraphics[scale=0.37]{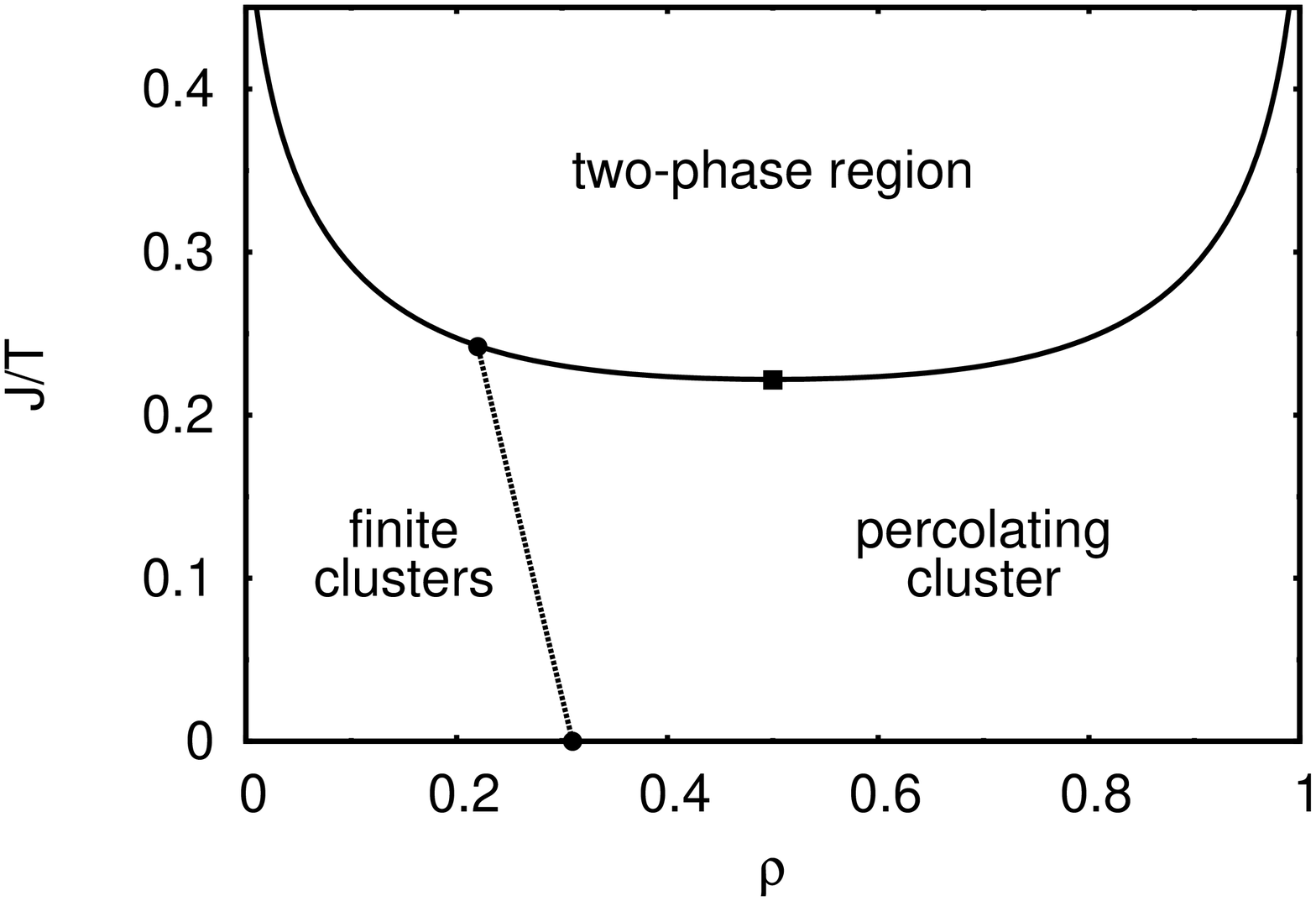}
\includegraphics[scale=0.37]{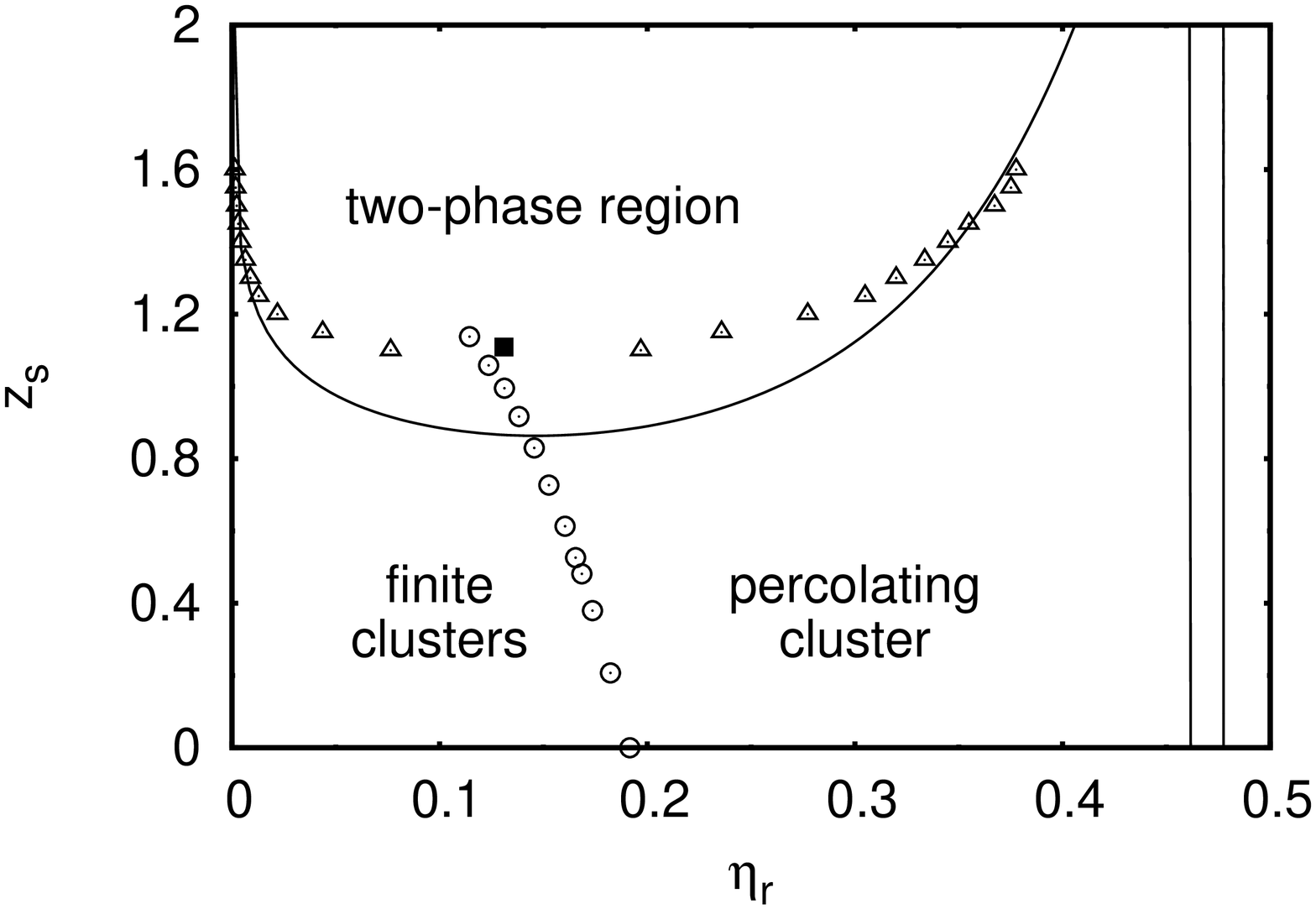}
\caption{\label{fig3} Schematic phase diagram of the lattice gas model (top), plotting inverse temperature $J/T$ ($J$ is the exchange constant of the corresponding Ising model) versus density $\rho$, and phase diagram of the rod-sphere mixture, for $L/D=3$, plotting fugacity $z_s$ versus rod packing fraction $\eta_r$ (bottom). In the lattice gas model, the percolation line moves from $\rho^{\rm perc} \approx 0.31$ for $J/T=0$ to $\rho^{\rm perc}\approx 0.22$ at the coexistence curve. For the rod-sphere mixture, results from free volume theory \cite{28} are included (full curves), while the symbols denote Monte Carlo data. Using an intensive variable at the ordinate, critical points appear as minima of the coexistence curve, unlike Fig.~1. The two vertical lines in the right part show the isotropic-nematic coexistence region \cite{28}.}
\end{figure}

Both, for the lattice gas model and the present rod-sphere model this variation of the percolation transition line has interesting consequences for the cluster morphology in the initial stage of spinodal decomposition \cite{37,38}: Performing a quenching experiment by suddenly enhancing $J/T$ or $z_s$, one may bring a system from the non-percolating state in the one-phase region to a state slightly inside the coexistence curve, where percolation should occur: then one should observe a kind of ``gelation transition'' during the initial stages of phase separation, where a percolating cluster forms rather quickly\cite{37,38}. However, during the coarsening stage it is often observed (e.g.~\cite{20}) that clusters compactify, and the initially percolating morphology breaks up into an assembly of many separate droplets. This remark illustrates the well-known fact \cite{39} that one must not associate the percolation transition between the interconnected morphology of a phase-separating system and the droplet morphology with the mean-field concept of a ``spinodal curve'' \cite{40}, where the free energy barrier against homogeneous nucleation is predicted to vanish, and which hence is considered as a dividing line between distinct mechanisms of phase separation kinetics, nucleation and growth on the one side, spinodal decomposition on the other side \cite{38,39,40}. Thus, assemblies of growing droplets can be the result of a spinodal decomposition process. We expect that these general considerations will carry over to phase separation kinetics of rod-coil mixtures, too.

A specific aspect of the latter systems, however, is the question how the percolation threshold depends on the aspect ratio $L/D$ \cite{24,41,42}. For ideal, non-interacting rods (which may overlap each other) one can show that $\eta_r^{\rm perc} \propto D/L$ for large $L/D$. However, for the present model such a behavior is not yet reached even for $L/D = 50$, rather $\eta_r^{\rm perc}$ decreases much more weakly with increasing $L/D$ \cite{24}. However, calculations based on the reference interaction site model (RISM) predict \cite{42} that the effect of the hard core interaction is to postpone the scaling with $L/D$ to larger aspect ratios, so $1/L$ scaling is eventually recovered. Nevertheless, a consequence of the observed decrease of $\eta_r^{\rm perc}$ with $L/D$ is that orientational correlations never are important near the percolation threshold. Another intriguing question concerns the issue whether the percolation exponents change when $L/D$ is varied \cite{41}. We suggest from the universality principle that different exponents possibly apply only in the limit $L/D \rightarrow \infty$, but for finite and large $L/D$ the asymptotic critical region is rather narrow. These questions are not of purely academic interest, since a very small volume fraction of long carbon nanotubes in a polymer matrix may hence suffice to provide a material with useful electrical conductivity \cite{24,41}.

\section{Percolation Versus Orientational Ordering in Systems of Hard Platelets}
Here, we consider platelets formed by cutting off two sphere caps of equal size from a hard sphere of diameter $D$, such that a disk of height
$L$ remains \cite{29}. Again, we define two disks as connected if the minimal distance between points at their surfaces is less then some (arbitrarily chosen) distance $A$ (mostly again $A=0.2 L$ is used). Again, the percolation threshold is found here from finite size scaling analyses of the percolation probability $\Pi(\eta)$ where $\eta$ now denotes the packing fraction of the disks (the volume of one disk is $v_d=\pi L(D^2-L^2/3)/4$, and $\eta =N_dv_d/V$ with $N_d$ the number of disks in the volume $V$). In this case, simulations were done only in the canonic ensemble (fixed particle number $N_d$), considering single-particle Monte Carlo moves consisting of small translations and rotations of the particles. For equilibration of the configurations, $1.5 \times 10^6$ trial moves per particle were used, and $3 \times 10^6$ moves for the sampling of the percolation properties. Just as in the case of rods, a finite size scaling analysis of $\Pi(\eta)$ was performed, studying various values of the aspect ratio $D/L$, as well as of the ratio $A/L$ defining our connectedness criterion. We also have paid attention to the occurrence of orientational order, by recording the nematic order parameter \cite{18}. If $\vec{u}_i$ is a unit vector perpendicular to the planar surface of a platelet (labeled by index i), then $S$ is given by the largest eigenvalue of the tensor $Q_{\alpha \beta}=(1/2 N_d) \sum \limits _i (3 u_i^\alpha u_i^\beta - \delta_ {\alpha \beta})$, as usual. Studying systems in the range from $N_d=900 $ to $N_d = 19200$, $\eta$ is varied by changing $V$, and $\eta ^{\rm perc}$ can be found very precisely, confirming again that the critical exponents are those of the standard percolation problem \cite{29}. For the sake of saving space, we direct the reader to a recent original publication \cite{29} for more details.

\begin{figure}
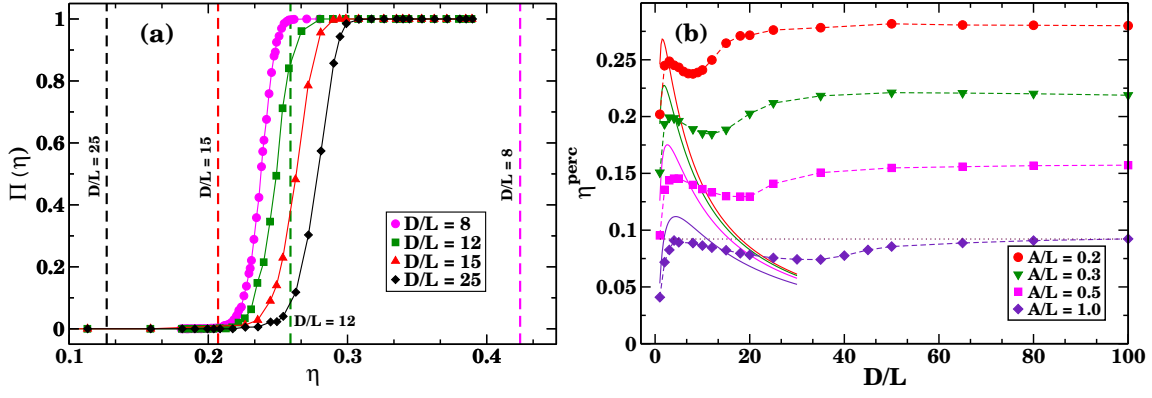

\centering
\includegraphics[scale=0.20]{PlateletsGraph1.eps}
\includegraphics[scale=0.20]{PlateletsGraph2.eps}
\caption{\label{fig4} (a) Percolation probability $\Pi(\eta)$ as a function of the platelet packing fraction $\eta$, for $A=0.2L$ and four values of $D/L$, as indicated, using $N_d=900$ platelets. The dashed vertical lines indicate the volume fractions where the nematic order parameter is $S=0.5$, yielding an approximate indication where for the shown choices of $D/L$ the isotropic nematic transition occurs. (b) Percolation threshold $\eta^{\rm perc}$ as a function of the aspect ratio $D/L$ of the disks, for four connectivity distances $A/L$. The solid lines represent theoretical predictions, resulting from a contact volume argument which would imply $\eta^{\rm perc}\approx L/D$ for large $D/L$ \cite{29}. The dotted horizontal line is a theoretical prediction from Ref.~\cite{43}. The dashed curves are guides to the eye only. From \cite{29}. }
\end{figure}

While for this model the dependence of $\eta^{\rm perc}$ on $D/L$ is relatively weak, the location of the threshold value where nematic order sets in depends on $\eta$ very strongly (Fig.~4a); note that our Monte Carlo simulations have not been precise enough to resolve the widths of the expected isotropic-nematic coexistence regime, which presumably is very narrow.

The fact that for large $D/L$ the system is nematically ordered has various interesting consequences for percolation. First of all, the existence of the director along which the platelets are aligned creates an anisotropy in the system, with the effect that the number of clusters that percolate along the director is smaller than in the directions perpendicular to it. Secondly, the contact volume predictions (assuming random orientation of the disks and leading to $\eta^{\rm perc} \propto L/D$ for large $D/L$) fail dramatically (Fig.~4a). Rather $\eta^{\rm perc}$ settles down at a plateau value for large $D/L$. Fig.~4b demonstrates that the actual value of the plateau must depend on the arbitrary parameter $A/L$ used to define connectivity, but the qualitative behavior is the same for all $A/L \leq 1$. One can understand this behavior qualitatively from a cell model argument, which states that each disk occupies an effective volume $V_{\rm eff}= \pi/4 D^2 L_{\rm eff}$, where $L_{\rm eff}= L +\beta D(1-S)^{1/2}$, with a parameter $\beta \approx 0.49$ adjusted to the actual simulation data. The enhancement of $L_{\rm eff}$ relative to $L$ is attributed to fluctuations of the disks around the director orientation, for the regime where $1-S \ll 1$. The volume of all disks then fits the available space completely. Arguing then that $A\propto L_{\rm eff}-L$, one indeed predicts that $\eta^{\rm perc}$ depends on $A/L$ but not on $D/L$, when $1-S\ll 1$ \cite{29}.

However, for the possible application of graphene platelets in electrically conducting composites the fact that $\eta^{\rm perc}$ does not decrease when $D/L$ gets large is unfavorable, because it means one will need relatively large amounts of graphene in order to have materials with good conductivity.

\section{Effects of hydrodynamic interactions on crystal nucleation in colloidal suspensions}
We already have emphasized that colloids are popular model systems to study cooperative phenomena (in particular, phase transitions) in condensed matter. The large size and the slowness of the dynamics of colloidal particles are favorable for experimental studies, and the effective interactions often are well approximated by very simple models, such as hard rods or hard spheres.

A particularly interesting problem which has found attention both from experiment \cite{44,45,45a} and from simulations \cite{46,47,48} is the nucleation of crystals of hard-sphere like colloids from the fluid phase. As far as static properties (e.g. the radial pair distribution function at various packing fractions in the fluid, etc.) of the considered sterically stabilized polymethacrylate particles (or polystyrene spheres in water, etc.) are concerned, one expects that simple hard spheres should be a very good model. However, the variation of the observed nucleation rates with the supersaturation of the colloidal suspension differs strongly from the corresponding simulation results \cite{44,45,45a,46,47,48}. The physical reason for this discrepancy hitherto has not been understood. Clearly, if such basic dynamic aspects of an archetypical system such as hard sphere colloids are far from being understood, it is doubtful that one can clarify the dynamics of anisotropic colloids, where a coupling between translational and rotational motions of the particles can be expected.

Thus we have opted to reconsider the kinetics of crystal formation for hard sphere colloids, but unlike early work \cite{46} where nucleation rates were estimated by umbrella sampling of the free energy barrier we shall address the actual nucleation kinetics of the system, paying attention to the role of the solvent. (The approach by means of umbrella sampling relies on the general theory of nucleation phenomena \cite{40}, thus the kinetics of the nucleation process is not explicitly considered there).

While in a molecular system, where all particles have similar sizes and masses, nucleation kinetics would be studied by Molecular Dynamics (MD) methods \cite{49}, it is impossible to apply this approach in a naive way to a colloidal system, due to the enormous disparity in size and mass between solvent molecules and colloidal particles. Often the effect of the collisions between solvent molecules and colloidal particles simply is described in terms of friction (plus random forces, i.e. one carries out a ``Brownian Dynamics'' simulation \cite{50}). But this approach misses the hydrodynamic interactions caused by solvent backflow. Brute force MD for asymmetric binary mixtures works at best for a difference in masses by a factor of about one order of magnitude \cite{51}, which is very far from the experiment. However, an elegant way out of this dilemma is provided by the multiparticle collision dynamics (MPC) method \cite{52,53,54,54a,54b}. In this method, the solvent fluid is described by an assembly of particles of mass $m$, which transport momentum through the system, respecting the conservation laws locally. The algorithm consists of two steps, free streaming of the fluid particles (where they move ballistically with their velocities) and multiparticle collisions, where these velocities change their orientation. This algorithm is combined with an event driven Molecular Dynamics method \cite{55,56,57} for the colloids which are modelled as hard spheres of diameter $a$ (see \cite{58} for details). By suitable choice of the density of these fluid ``pseudo-particles'' and their mass the solvent viscosity can be varied (the validity of this approach can be tested and the viscosity can be measured by setting up a simulation of Poiseuille flow confining the fluid between parallel plates with stick boundary conditions \cite{58}).

\begin{figure}
\includegraphics[scale=0.37]{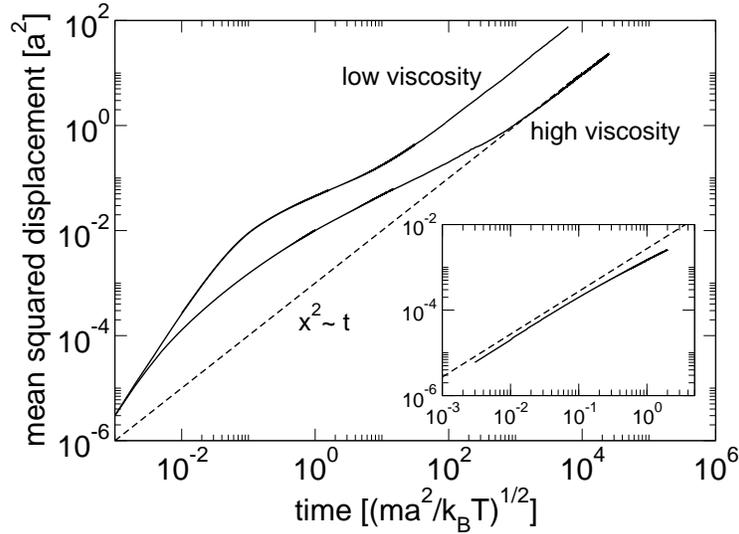}
\caption{\label{fig5} Mean squared displacement of a particle in the suspension versus time for a viscosity $\eta = 4.17 (mk_BT)^{1/2}/a^2$ and $\eta = 63.93 (mk_BT)^{1/2}/a^2$ (main panel) and 915 $\sqrt{mk_BT}/a^2$ (inset). From \cite{58}}
\end{figure}

\begin{figure}
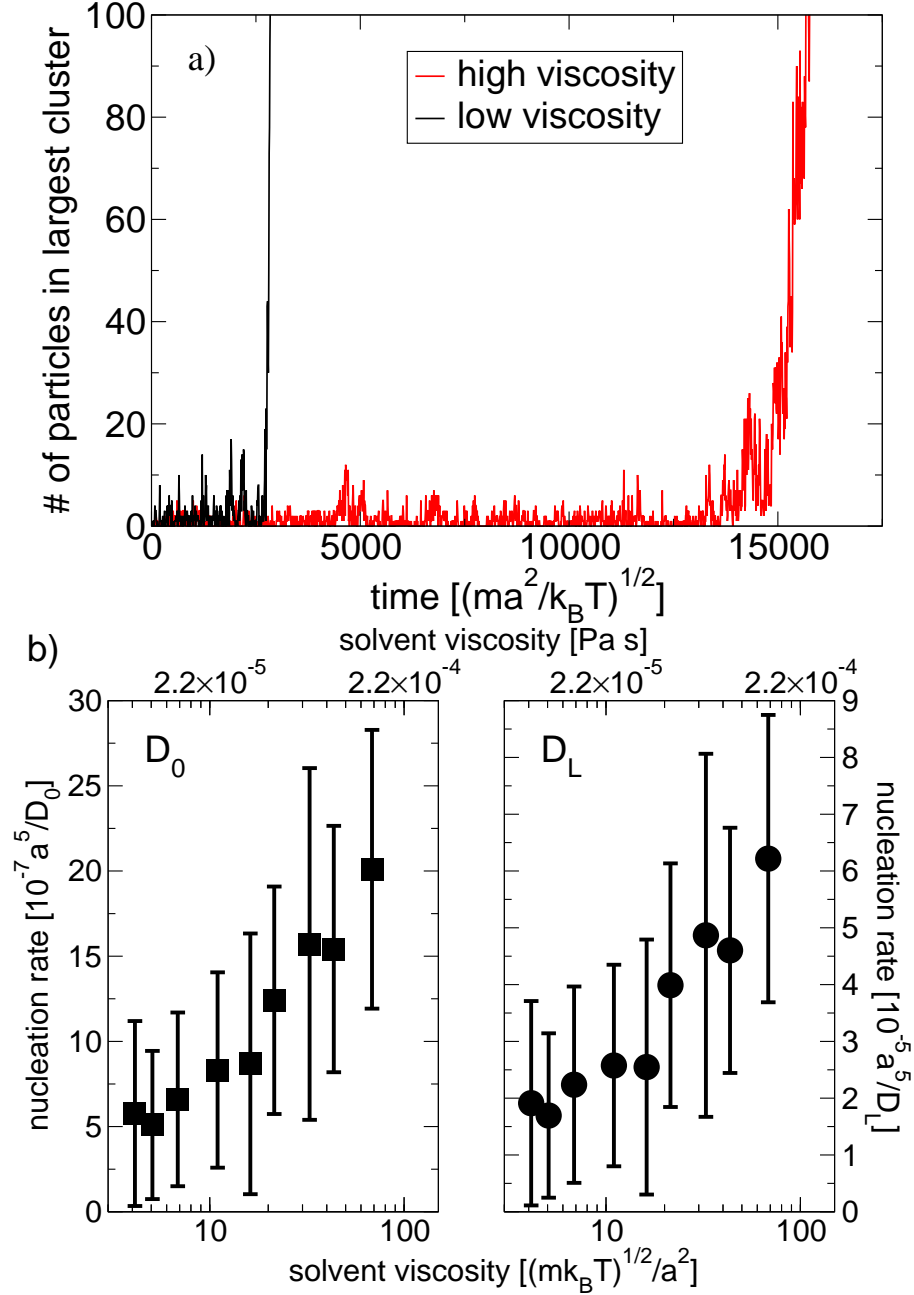

\includegraphics[scale=0.3]{HSTimeseries.eps}
\includegraphics[scale=0.3]{HS_NucleationRates.eps}
\caption{\label{fig6} (a) Number of particles in the largest cluster as a function of time for two trajectories, one for a low value of the viscosity ($\eta = 4.17 (mk_BT)^{1/2}/a^2$) and the other for a high value ($\eta = 63.93 (mk_BT)^{1/2}/a^2$). From \cite{58} (b) Nucleation rate scaled by $D_0$ (left panel) and $D_L$ (right panel) as a function of solvent viscosity. From \cite{58}}
\end{figure}

The first task then is to obtain the diffusion constant of the colloidal particles, since this sets the time scale allowing a comparison with experiment. This diffusion constant is extracted from following the mean square displacement of the particles with time (Fig.~5). One sees that the viscosity $\eta$ can be varied over more than two orders of magnitude (the largest choice shown would correspond to $2\cdot 10^{-3}$ Pa$\cdot$s for particles with a diameter of 420 nm, which is of the same order as reached in experiments \cite{1,2,3}).

Typical simulations used 8240 hard spheres at packing fraction $\eta = 0.539$ and $\eta = 0.544$, well above the onset of crystallization $(\eta_f=0.494$). The starting configurations were prepared in the supersaturated liquid, and using the standard $q^6$-bond order parameter \cite{59,60} it was verified that initially no crystalline nuclei were present. To monitor nucleation events, the number of particles in the largest crystalline cluster was monitored as a function of time (Fig.~6a). One sees that this number fluctuates below 30, but when it exceeds 30 it rapidly grows to a large size (in at least 50\% of the cases). This is the typical picture of nucleation, a rare event after some time lag. One also sees that nucleation takes longer for high solvent viscosity. The nucleation rate then is obtained sampling this induction time $t_i$ that one needs to wait before nucleation events occur, $I=(V\langle t_i\rangle )^{-1}$, $V$ being the system's volume. (This relation between the average induction time and the nucleation rate should hold for our simulations, because the system is relatively small. We observe only one nucleation event per simulated trajectory which results in an immediate crystallization of the entire system without interference of other nucleation events.) Fig.~6b shows the nucleation rate as a function of solvent viscosity, scaled either by the diffusion constant $D_0$ of a particle in the dilute limit (left), or by the actual long time diffusion constant $D_L$ in the suspension (right part). There is a clear increase of the rate with viscosity when it is scaled in this way. Thus nucleation rates measured in suspensions with different solvents cannot be straightforwardly compared, crystallization kinetics is affected by solvent hydrodynamics substantially. This shows already that the previous comparison of simulations \cite{46,47,48} where hydrodynamics was ignored, with experiments (where it is inevitably present) presumably are not so meaningful. We note that also in phase separation kinetics of colloid-polymer mixtures the importance of hydrodynamic interactions could be demonstrated by simulations using the MPC method \cite{20}.

\section{Crystallization Kinetics of Hard Ellipsoids Close to the Glass Transition}
While in the previous section it was demonstrated that hydrodynamics is important for the crystallization of perfectly spherical hard spheres, where crystallization happens via nucleation followed by subsequent rather rapid growth (Fig.~6a), in systems of hard ellipsoids crystallization is more difficult, and then one can bring the disordered fluid deeper into the thermodynamically unstable region, where crystallization proceeds without the need of crossing high free energy barriers by nucleation events, but is nevertheless slow because such a strongly overcompressed system is close to its glass transition. For weaker overcompression, one can also observe the standard nucleation and growth scenario \cite{21}, similar to the hard sphere case just discussed, but this is out of our scope here. For systems of monodisperse ellipsoids with aspect ratio $a/b = 1.25$, the equilibrium phase diagram has been established \cite{61,62,63} and also the presence of a glass transition has been demonstrated \cite{64}. For dense ellipsoids, hydrodynamics of the solvent clearly does not matter, and hence on a qualitative level the dynamics of a dense system of ellipsoids can be very well modeled simply by single particle Monte Carlo (MC) moves \cite{21}. These moves consisted of random displacements of the center of mass of the particles (up to 0.03 particle diameters for elementrary MC step) and random rotations of the particle axis (up to 1.8$^o$). Simulations are carried out in the constant particle number ($N$) and constant pressure ($P$) ensemble, for $N=10386$ particles. It is known that liquid-solid coexistence occurs for $P^*=8abP/k_BT=14.34$ \cite{63a}, and choosing $P^*=27$ to $P^*=50$ the fluid clearly is out of equilibrium. While for $P^*\leq 30$ the time-dependence of the mean square displacements indicates that still ordinary diffusion occurs (Fig.~7a), for $P^* \geq 40$ there is clear evidence of sub-diffusive behavior, and also the dynamic structure factor (Fig.~7b) decays according to the stretched exponential behavior that is typical for glass forming systems \cite{30}. While one finds that for moderate overcompression $(P^*=27)$ the crystallization kinetics still is similar to the standard nucleation mechanism of the previous section, for strong overcompression $(P^*=40$) the situation is different, crystallization sets in immediately, and huge crystalline regions form without the need of substantial diffusion of the particles. The crystal phase is found to rapidly form a percolating network, despite the approach to glassy dynamics \cite{21}. In this context it is natural to ask whether the specific aspects of glassy dynamics then affect crystallization. For instance, one very important signature of glassy dynamics is ``dynamic heterogeneity'' \cite{30}: a fluid close to the glass transitions exhibits regions where the particles either move orders of magnitude more slowly or faster than on average. Thus, one might expect that crystallization occurs preferentially in the regions of the more mobile particles. However, although it seems rather natural to assume that the necessary structural rearrangements in order to transform a fluid region into a crystalline region are facilitated by higher mobility of the particles, the simulations clearly show the absence of any correlation between dynamic heterogeneities and crystallization (Fig.~8).

\begin{figure}
\includegraphics[scale=0.3]{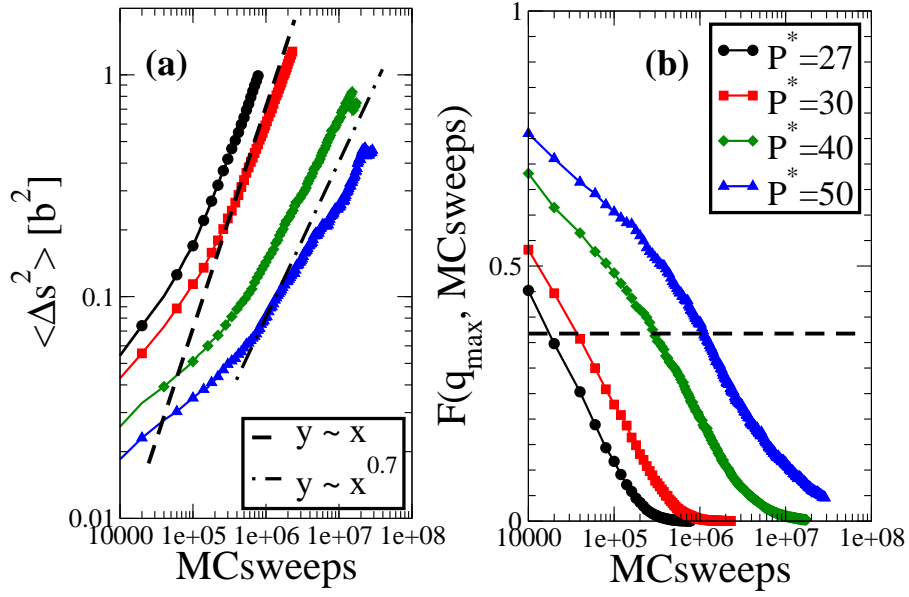}
\caption{\label{fig7} (a) Mean square displacements $\langle \Delta s^2 \rangle$ of hard ellipsoids as a function of  simulation time (measured in the units of Monte Carlo (MC) sweeps)) for different values of the pressure $P^*$ (indicated in (b)). Dashed and dash-dotted straight lines on this log-log plot indicate ordinary diffusion ($\langle \Delta s^2\rangle \propto t$) and sub-diffusive behavior $(\langle \Delta s^2 \rangle \propto t^{0.7})$, respectively. (b) Incoherent dynamic structure factor $F(q_{\rm max},t)$ for wave-numbers $q$ chosen at the maximum $q_{\rm max}$ of the coherent static structure factor versus time, for four pressures $P^*$ as indicated. From Dorosz and Schilling \cite{21}.}
\end{figure}

\begin{figure}
\includegraphics[scale=0.3]{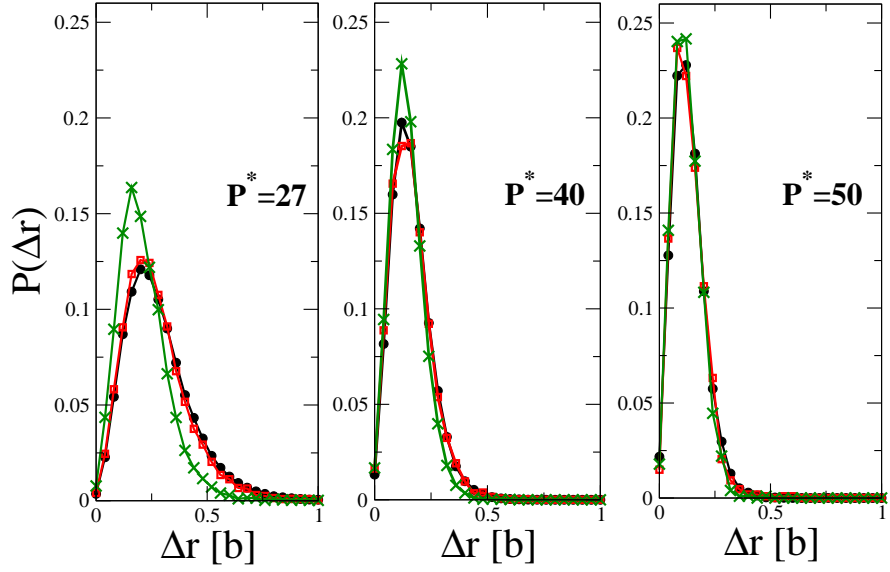}
\caption{\label{fig8} Mobility in the surrounding liquid (circles) in comparison to the mobility of particles that are going to crystallize (squares) and of the particles at the surface of the crystal, once the crystallite has formed (crosses). ``Mobility'' here is measured in terms of the probability distribution of the distance that a particle has traveled during a time interval of $5\cdot 10^4$ MC sweeps. From \cite{21}.}
\end{figure}

\section{Concluding Remarks}
In this paper, some aspects of the statistical mechanics of simple models for colloidal suspensions were discussed, such as hard rods, hard ellipsoids or even simple hard spheres, as well as systems containing hard platelets. One aspect that was discussed in detail was the possible interplay of the formation of a percolating network of colloidal particles with equilibrium phase transitions, such as orientational ordering into nematic phases, or phase separation into vapor-like and liquid-like phases (driven by attractive depletion interaction among the colloids, when polymers are present in the solution). We have demonstrated that the percolation transition is characterized by the critical exponents of standard random percolation, irrespective of all the correlations between the colloidal particles. In particular, the aspect ratio of the particles affects the location of the percolation threshold only, but not the exponents. While increasing aspect ratio causes a decrease of the percolation threshold for rods, this is not the case for platelets, and this fact can be understood as a consequence of the pronounced nematic order.

Also aspects of the dynamics of colloids in the fluid state and its effect on crystallization kinetics have been investigated. For hard ellipsoids, it is feasible to study a spinodal-decomposition like crystallization of highly overcompressed fluids, but the glassy single-particle dynamics does not have any specific consequences on crystallization, apart from setting the overall time-scale. In contrast, for the simple hard-sphere system it was found that varying solvent viscosity has a pronounced effect on the nucleation rate of crystals. We expect that solvent-mediated hydrodynamic interactions should always be included, when one deals with the cooperative dynamics of colloidal suspensions at moderate densities.

\section*{Acknowledgements}
This work was supported by the DFG within SFB TR6 (project D5). Marc Radu acknowledges financial support by the National Research Fund, Luxembourg, AFR scheme (PHD-09-177). Sven Dorosz acknowledges financial support by the National Research Fund, Luxembourg, co-funded under the Marie Curie Actions of the European Commission (FP7-COFUND) and the National Research Fund Luxembourg under the project FRPTECD. Computer simulations presented in this paper were carried out using the HPC facility of the University of Luxembourg.
We are grateful to M. Miller, M. Oettel, and R. Tuinier for fruitful collaboration on some aspects of this work.

%References
%\bibliography{codef_refs}
%\bibliography{bookrefs_merge}
%\bibliography{references}

\end{document}